\documentclass[12pt]{article}
\usepackage{graphicx}

\newsavebox{\hflrar}
\sbox{\hflrar}{\makebox[0pt][l]
{${\scriptstyle \leftharpoonup}$}{${\scriptstyle \rightharpoonup}$}}

\def \jpsi {J/\psi}

\def \to {\rightarrow}

\def \jpsi {J/\psi}

\def\FT{T_R \vert_{t\to 0}}

\begin{document}
\begin{flushright}
AS-ITP-2002-002
\end{flushright}
\pagestyle{plain}
\vskip 10mm
\begin{center}
{\Large Ratio of Photoproduction Rates of $\jpsi$ and $\psi(2S)$ } \\
\vskip 10mm
J. P. Ma   \\
{\small {\it Institute of Theoretical Physics , Academia
Sinica, Beijing 100080, China }} \\
~~~ \\
Jia-Sheng Xu \\
{\small {\it China Center of Advance Science and Technology
(World Laboratory), Beijing 100080, China }} \\
{\small {\it and Institute of Theoretical Physics , Academia
Sinica, Beijing 100080, China }}
\end{center}

\vskip 0.4 cm


\begin{abstract}
There are different approaches for diffractive photoproduction of charmonia.
Recently, a new approach is proposed, in which charm quarks are taken
as heavy quarks and the nonperturbative effect related to charmonia can be
handled with nonrelativistic QCD. The interaction between the $c\bar c$ pair
and the initial hadron is through exchange of soft gluons. The exchange of soft
gluons can be studied with heavy quark effective theory and an expansion
in the inverse of charm quark mass $m_c$ can be employed. In this approach a simple
formula for the S-matrix can be derived by neglecting higher orders in $m_c^{-1}$
and relativistic correction related to charmonia. The S-matrix is related
to the usual gluon distribution $g(x)$ at small $x$. This result is different
than those from other approaches. Confronting experiment the result is not
in agreement with experimental measurement because large errors from higher
order in $m_c^{-1}$ and from relativistic corrections. Nevertheless  the ratio of cross
sections of $\jpsi$ and $\psi(2S)$ can be predicted more precisely than cross-sections.
In this letter we show that the ratio predicted in this approach with an estimation
of relativistic corrections is in good agreement with the recent measurement at HERA.

\end{abstract}

\vfill\eject\pagestyle{plain}\setcounter{page}{1}



Diffractive photoproduction of a charmonium has been studied extensively.
Experimentally diffractive photoproduction of $\jpsi$ and $\psi(2S)$ has been
measured with increasing precision at HERA\cite{Herajpsi, Herapsi,psinew},
while there are different approaches for the process\cite{MR,FS,Nem,HKT,SHIAH,maxuj}.
Recently H1 has studied diffractive photoproduction of $\psi(2S)$ and obtained
the ratio\cite{psinew}
\begin{equation}
 R=\frac {\sigma(\psi(2S))}{\sigma({\jpsi})} =0.166\pm 0.007(stat.)
   \pm 0.008(sys.) \pm 0.007(BR),
\end{equation}
which is in agreement with an earlier measurement by H1\cite{Herapsi}.
The measured ratio has a weak dependence on the invariant mass $s$
of the initial states.
Although this ratio  and the dependence on $s$ can be explained theoretically,
e.g., with the approaches presented in \cite{Nem,HKT}, but
there are always certain free parameters in these approaches. In this letter we will show
that the ratio and the weak $s$-dependence can be well explained
with the soft gluon approach proposed in \cite{maxuj}, in which relativistic correction
for charmonia is taken into account and predictions are made without free parameters.
\par
In previous approaches\cite{FS,Nem,HKT,SHIAH}
the nonperturbative properties of charmonium is described by light-cone wave-functions,
which are defined with operators classified by twists. For diffractive photoproduction
only a leading-twist wave-function is used. It is then questionable if contributions
from higher-twist wave-functions are suppressed or not, and also for consistency
of the twist-expansion the c-quark should be taken as a light quark and its mass
$m_c$ should be neglected. However, taking $m_c=0$ will cause some problems
in these approaches and hence $m_c$ is not neglected. Because
the leading-twist wave-function is unknown, either a parameterization is
introduced or it is related to the nonrelativistic wave-function of charmoina.
By solving Schr\"odinger equation with a suitable potential,
the nonrelativistic wave-function, then the leading-twist wave-function is
obtained through a Lorentz boost. It is clear that certain free parameters are introduced for
determining the leading-twist wave-function. For the interaction between
the initial hadron and the $c\bar c$-pair one usually uses the small-size
color dipole approximation, in which some modification is also introduced,
e.g., in  \cite{FS}. Hence the color dipole approximation also contains
some parameters, which need to be determined from other sources.
\par
A new approach\cite{maxuj}, which is different than the above discussed, is used
to study diffractive photoproduction of $\jpsi$. In this approach the $c$-quark is
taken as a heavy quark, hence one can use nonrelativistic QCD(NRQCD) to describe
the nonperturbative property of charmonia\cite{nrqcd} by noting the fact
that the $c$-quark in $\jpsi$ in its
rest-frame moves with a small velocity $v$. A systematic expansion
in $v$ can be employed. At leading order of $v$ the chamonium can be
taken as a bound state of a $c\bar c$ pair. The interaction between
the $c\bar c$ pair and the initial hadron is through exchange of gluons.
These gluons are soft in the diffractive region. Because the c-quark is
taken as a heavy quark, the heavy quark effective theory(HQET)\cite{HQET}
can be used to study the exchange of soft gluons and an expansion
in $m_c^{-1}$ for the exchange can be performed systematically. This approach
was first used to study the decay $\jpsi\to e^+e^- +\pi+\pi$ where the
two pions are soft\cite{maxu}. Keeping only leading orders, results for
the S-matrix element are obtained and they are different than
those in \cite{MR, FS}. The reason for this is discussed in detail in \cite{maxuj}.
\par
Considering the process
\begin{equation}
\gamma + p\to p +\jpsi
\end{equation}
in the diffractive region, the differential cross-section in the approach
in \cite{maxuj} is given by:
\begin{equation}
 \frac{d\sigma}{dt}\vert_{t\to 0} = \frac{4}{9}\frac{Q_c^2 \alpha_em} {s^2m_c^5}
 \cdot \langle \jpsi\vert O_1(^3S_1)\vert \jpsi\rangle\cdot
  \overline{\sum} \vert T_R \vert ^2\cdot\left\{
    1 +{\cal O}(\frac{\Lambda_{QCD}}{m_c}) +{\cal O}(v^2)\right\},
\end{equation}
where $\overline{\sum}$ is the summation over spin of the final hadron $h$ and
the spin average of the initial hadron. $t$ is the squared momentum transfer
between protons. The matrix element
$\langle \jpsi\vert O_1(^3S_1)\vert \jpsi\rangle$ is defined with NRQCD
fields\cite{nrqcd} and is related to the leptonic decay width
\begin{equation}
\Gamma (J/\psi\to e^+e^- ) = \alpha_{em}^2 Q_c^2 \frac {2\pi}{3m_c^2}
  \cdot\langle \jpsi\vert O_1(^3S_1)\vert \jpsi\rangle
  \cdot\left\{
    1 +{\cal O}(v^2)\right\}.
\end{equation}
The nonperturbative effect of the initial proton is represented by $T_R$ and $T_R$
is defined by a matrix element of field
strength operators, which are separated in the moving direction of
$J/\psi$ in the space-time. In the case $s\gg m_c$, the contribution
from leading-twist operator is dominant and $T_R$ can be related
to the usual gluon distribution $g(x)$ by:
\begin{equation}
\FT \approx -i\pi^2m_c \alpha_s^2(m_c) g(x_c) \left[ 1
  - i \tan(\frac{1}{2}\alpha\pi) \right],
\end{equation}
with
\begin{equation}
  x_c= \frac {2m_c^2}{s}.
\end{equation}
The term with $\tan(\frac{1}{2}\alpha\pi)$ is obtained by assuming
$xg(x)\to x^{-\alpha}$ for $x\to 0$. In Eq.(5) we have set the renormalization
scale  $\mu$ to be $m_c$. The total cross section
is obtained by:
\begin{equation}
  \frac{d\sigma(\jpsi)}{dt} = \frac{d\sigma(\jpsi)}{dt}\vert_{t\to 0}
  \cdot \exp^{-b_{\jpsi} \vert t\vert},
\end{equation}
where the slope parameter $b$ is measured in experiment for $\jpsi$ and
for $\psi(2S)$.
\par
It should be noted that at the orders
we consider $\jpsi$ has the same helicity as that of the initial photon and
the final proton also has the same helicity as that of the initial one.
In Eq.(3) and (4) we also give the orders of theoretical errors.
One of errors is from the expansion in $m_c^{-1}$ for the exchange
of soft gluons, which is at the order of $\lambda_{QCD}/m_c$. The parameter
$\lambda_{QCD}$ is the typical scale of nonperturbative QCD and is at order
of several hundreds MeV. Another is the relativistic correction at order
of $v^2$, because we only used leading order in the expansion in $v$ for
$\jpsi$. These errors are likely very large and result in that the predicted
cross-section of $\jpsi$ is not in agreement with experiment. For $\Upsilon$,
these errors are significantly smaller than those for $\jpsi$, and the
predicted cross-section of $\Upsilon$ agrees with experiment fairly well\cite{maxuj}.
Although these errors are large for charmonia, but some of them is cancelled
in the ratio defined in Eq.(1), e.g., the error at order of $\lambda_{QCD}/m_c$.
Using the above results we obtain the prediction:
\begin{equation}
 R= \frac{b_{\jpsi}}{b_{\psi(2S)}}\cdot \frac {\Gamma (\psi(2S) \to e^+e^-)}
             {\Gamma (\jpsi \to e^+e^-)}\cdot\left\{
             1+{\cal O}(v^2)\right\}.
\end{equation}
In this ratio some corrections at higher orders of $\alpha_s$ are also cancelled.
Using the experimentally measured slop parameters
$b_{\jpsi} =(4.99\pm 0.13\pm 0.39)$GeV$^2$ and
$b_{\psi(2S)}=(4.31\pm 0.57\pm0.46)$GeV$^2$, and the experimental data for the
leptonic decay widths we obtain:
\begin{equation}
 R\approx 0.47.
\end{equation}
\begin{figure}[hbt]
\centering

\includegraphics[width=130mm,height=91mm]{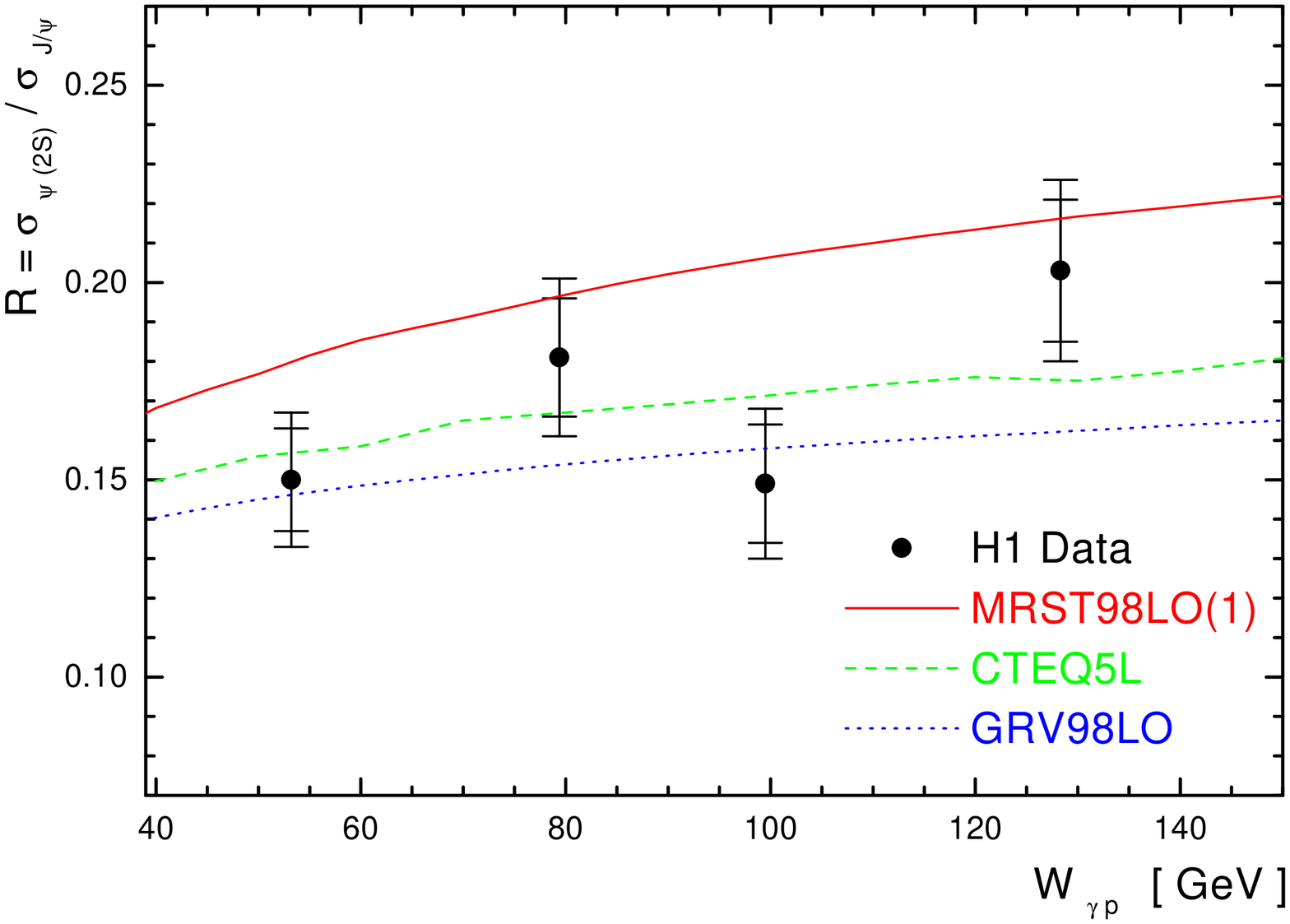} 
\vspace{-10mm} 
\caption{Prediction of Eq.(12) for the ratio R,
where the cures are drawn with different parameterizations of the
gluon distribution. Experimental data are from \cite{psinew}. }
\label{Feynman-dg1}
\end{figure}
This is almost three times of the experimentally measured. With the prediction
the ratio in Eq.(1) should have no $s$-dependence.
Experimentally the dependence is observed and
is parameterized by $R\propto s^\delta$ with $\delta=0.24\pm0.17$\cite{psinew}.
This is also in conflict with our prediction.
\par
It is clear that the relativistic correction plays an important role
for predicting the ratio. The correction is unknown and requires
to be analyzed. The analysis for this correction will be in general
complicated. Nevertheless, in the framework of NRQCD\cite{nrqcd}
the differential cross-section can be written as:
\begin{equation}
\frac{d\sigma(\jpsi)}{dt }\vert_{t\to 0} =
     {\cal A}_1\cdot \langle \jpsi\vert O_1(^3S_1)\vert \jpsi\rangle
    +{\cal A}_2 \cdot \langle \jpsi\vert P_1(^3S_1)\vert \jpsi\rangle
       +{\cal O}(v^4),
\end{equation}
where ${\cal A}_1$ can be read from Eq.(3) and ${\cal A}_2$ is unknown.
The relativistic correction is characterized by the matrix element
$\langle \jpsi\vert P_1(^3S_1)\vert \jpsi\rangle$, whose definition
can be found in \cite{nrqcd}. With equation of motion it can be shown
\cite{GK}:
\begin{equation}
\langle \jpsi\vert P_1(^3S_1)\vert \jpsi\rangle =m_c(M_{\jpsi}-2m_c)
\cdot\langle \jpsi\vert O_1(^3S_1)\vert \jpsi\rangle\cdot
\left\{1+{\cal O}(v^2)\right\}.
\end{equation}
Replacing $\jpsi$ with $\psi(2S)$ one obtains the relation for
matrix elements of $\psi(2S)$. It should be noted that the pole
mass $m_c$ is not well determined yet, it lies in the range from
1.3GeV to 1.7GeV. Based on the relation in Eq.(11) an estimation
of relativistic corrections can be made if one takes $m_c$ is free parameter
and sets $2m_c=M_{\jpsi}$ for $\jpsi$ and $2m_c=M_{\psi(2S)}$ for $\psi(2S)$
respectively,
then the error due to relativistic correction at order $v^2$ is absent formally and
the correction is taken into account. With this setting it has been shown that
one could reduce the violation of the famous $14\%$ in the case of radiative decays
into $\eta'$ of $\jpsi$ and $\psi(2S)$\cite{Ma}. However,
it should be noted that this should be regarded as a phenomenological estimation,
a detailed analysis and a precise determination of $m_c$ is needed to
study the correction in a consistent way.
\par
Using the setting we obtain the ratio as:
\begin{equation}
 R= \frac{b_{\jpsi}}{b_{\psi(2S)}}\cdot \frac {\Gamma (\psi(2S) \to e^+e^-)}
             {\Gamma (\jpsi \to e^+e^-)}\cdot
             \frac {\alpha_s(\frac{1}{2} M_{\psi(2S)})}
                   {\alpha_s(\frac{1}{2} M_{\jpsi})}
\cdot \left (\frac {M_{\jpsi}}{M_{\psi(2S)}}\right)^5
\cdot \left [ \frac {x_2g(x_2)}{x_1g(x_1)}\right ]^2,
\end{equation}
with
\begin{equation}
  x_2=\frac {M^2_{\psi(2S)}}{2s},\ \ \ \ \
  x_1=\frac {M^2_{\jpsi}} {2s}.
\end{equation}
In Eq.(12)
we have neglected the terms with $\tan(\frac{1}{2}\alpha\pi)$. This term
has little effect on the ratio. With available
gluon distributions\cite{GRV,CTEQ,MRST98LO} we obtain numerical results
given in Fig.1, where the experimental data in \cite{psinew} is also drawn.
\par
From Fig.1 we can see that the prediction based on Eq.(12) agrees with
experimental data very well. Also, the weak $s$-dependence is predicted in which
$R$ increases with increasing $s$. This agreement shows clearly how important
the relativistic correction for charmonia is. From Eq.(12) one can see that the most
important factor is $(M_{\jpsi}/M_{\psi(2S)})^5=0.419$ for reducing the predicted
$R$-value in Eq.(9). Hence, the relativistic correction in $R$ can be larger  than
$50\%$. It should be noted that by taking
$2m_c=M_{\jpsi}$ for $\jpsi$ and $2m_c=M_{\psi(2S)}$ for $\psi(2S)$ respectively
the uncertainty from ${\cal O}(\lambda_{QCD}/m_c)$ in Eq.(3) will be not cancelled
in the ratio defined in Eq.(1), but it appears in the ratio at order of
$\lambda_{QCD}(M_{\psi(2S)}-M_{\jpsi})/(M_{\jpsi}M_{\psi(2S)})\approx 0.03$ by taking
$\lambda_{QCD}=500$MeV, hence the uncertainty is essentially smaller
in the ratio than that in the differential cross-sections.

\par
For $\Upsilon$-systems, if we neglect the relativistic correction and
correction from higher orders of $m_b^{-1}$,  we obtain the ratio
\begin{equation}
\frac {\sigma(\Upsilon(2S))}{\sigma({\Upsilon(1S)})} = \frac {\Gamma (\Upsilon(2S) \to e^+e^-)}
             {\Gamma (\Upsilon(1S) \to e^+e^-)}\approx 0.39,
\end{equation}
where we assume the same slope parameters for $\Upsilon(1S)$ and $\Upsilon(2S)$.
If we take the relativistic correction in analogy to Eq.(12) into account, i.e.,
by taking $2m_b=M_{\Upsilon(1S)}$
for $\Upsilon(1S)$ and $2m_b=M_{\Upsilon(2S)}$ for $\Upsilon(2S)$ respectively,
we obtain numerical results for
the ratio with available
gluon distributions\cite{GRV,CTEQ,MRST98LO}. The results are given in Fig.2. From
Fig.2 we can see that the predicted ratio has a significant deviation from
the one given in Eq.(14). Again, the most important factor here for reducing
the value given in Eq.(14) is $(M_{\Upsilon(1S)}/M_{\Upsilon(2S)})^5=0.75$.
This indicates that the relativistic correction in the ratio is also significant and
this significance may be beyond the expectation based on that the $b$-quark moves
inside a $\Upsilon$-system with a velocity $v^2\sim 0.1$. The reason for the significance
in the ratio may be explained by the following: The pole mass $m_b$ is known
more precisely than $m_c$. A lattice determination gives $m_b=5.0\pm 0.2$GeV\cite{mb}.
With this value
and experimental data of masses of $\Upsilon$-systems, one knows
that the matrix element of relativistic correction for $\Upsilon(1S)$ in Eq.(10)
is negative,
while for $\Upsilon(2S)$ it is positive, based on the similar relation given
in Eq.(11) for the case of $\jpsi$. It results in that the relativistic correction
in the ratio is the sum of absolute values of the both corrections.
Hence, the correction is significant. Experimentally, the ratio is not measured. If
it is measured, then our prediction given in Fig.2. can be tested.

\begin{figure}[hbt]
\centering
\includegraphics[width=130mm,height=91mm]{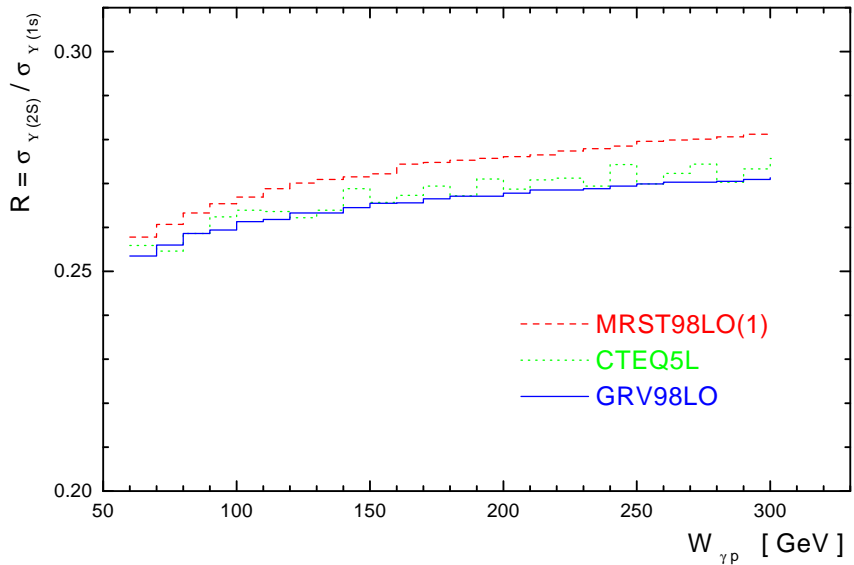} 
\vspace{-10mm} 
\caption{The Predicted production ratio for $\Upsilon$-systems,
where the cures are drawn with different parameterizations of the
gluon distribution. } \label{ratio-up}
\end{figure}

To summarize: In the soft gluon approach for diffractive photoproduction of $1^{--}$
charmonium we take
$c$-quark as a heavy quark, this allows one to use NRQCD to describe nonperturbative
properties of the quarkonium and to use HQET to handle the exchange of soft gluons
between the $c\bar c$-pair and the initial proton.
In this letter we use this approach to predict the ratio of the cross sections of
$\jpsi$ and $\psi(2S)$. It is shown that the prediction by adding relativistic correction
is in good agreement with the recently measured ratio. This indicates that
the relativistic correction is very important for charmonia. Finally, it should be noted
that our estimation
for relativistic correction is a phenomenological estimation based on theoretical
results, a detailed analysis and an precise determination of $m_c$ is needed to
study the correction in a consistent way.

\vskip 5mm
\begin{center}
{\bf\large Acknowledgments}
\end{center}

The work of J. P. Ma is supported  by National Nature
Science Foundation of P. R. China and by the
Hundred Young Scientist Program of Academia Sinica of P. R. China,
the work of J. S. Xu is supported by the Postdoctoral Foundation of P. R. China and  by
the K. C. Wong Education Foundation, Hong Kong.

\vfil\eject

\end{document}